\documentclass[12pt]{article}  

\usepackage{graphicx,color,epsf} 

\usepackage{amsmath,latexsym}

\begin{document}

\begin{center}
{\LARGE \bf 
Blackbody thermodynamics in the presence of Casimir's effect
}
\\ 
\vspace{1cm}
{\large  E. S. Moreira Jr.$^{a}$ and Heitor da Silva}$^{b,}$\footnote{E-mail:
{\tt moreira@unifei.edu.br} \& {\tt heitor@fisica.ufrn.br}}
\\
\vspace{0.3cm}
{$^{a}$\em Instituto de Matem\'{a}tica e Computa\c{c}\~{a}o,}
{\em Universidade Federal de Itajub\'{a},}   \\
{\em Itajub\'a, Minas Gerais 37500-903, Brazil}  \\
\vspace{0.1cm}
{$^{b}$\em Departamento de F\'{\i}sica Te\'{o}rica e Experimental,}
{\em Universidade Federal do Rio Grande do Norte,}   \\
{\em Natal, Rio Grande do Norte  59078-970, Brazil}

\vspace{0.3cm}
{\large February, 2023}
\end{center}
\vspace{0.5cm}


\abstract{ 
This paper is a study of the electromagnetic radiation at temperature $T$
in a rectangular slab whose walls are made of a perfect conductor. The two large parallel
walls of area $A$ are apart by a distance $d\ll \sqrt{A}$.
We take $T$, $A$, and $d$ as thermodynamic parameters, obtaining
the free energy from a procedure that involves the integration over the slab
of the 
ensemble average 
of the stress-energy-momentum tensor 
calculated long ago by Brown and Maclay. Both thermodynamic regimes
$kTd/\hbar c\gg 1$ and $kTd/\hbar c\ll 1$ are fully addressed.
We show that certain thermodynamic quantities which are notoriously 
ill defined (or trivial) in ordinary blackbody thermodynamics are now 
well defined (or nontrivial) due to presence of boundary conditions at the walls of the slab (``Casimir's effect").
The relationships among such quantities are fully explored and it is
speculated that they may be experimentally checked.
Stability is addressed, showing that electromagnetic radiation in the slab
is thermally stable; but mechanically unstable.
We investigate thermodynamic processes where
temperature, internal energy,  entropy and enthalpy
are each taken 
to be constant, revealing rather atypical behaviors.
For example,
in sharp contrast with what one would expect from a gas, when 
$kTd/\hbar c\ll 1$, ``free expansion" gives place to ``free contraction''
in accordance with the second law of thermodynamics. 
As a check of consistency of the 
formulae we remark that various Carnot cycles have been examined and verified that they correctly lead to Carnot's efficiency.

\vspace{0.5cm}

\section{Introduction}
\label{introduction}
Statistical mechanics is plenty of counter-intuitive phenomena. A well known is
the zero point pressure (pressure $P$ as temperature $T\rightarrow 0$)
of an ideal Fermi gas with $N$
molecules of mass $m$ in a volume $V$
\cite{hua87}. The corresponding equation of state is given by
(along the text, the fundamental constants $k$, $\hbar$ and $c$
have the usual meaning)
\begin{eqnarray}
&&P=\frac{2}{5}\frac{N}{V}\epsilon_{F}
\left[1+\frac{5\pi^{2}}{12}\left(\frac{kT}{\epsilon_{F}}\right)^{2}+\cdots\right],
\label{p0}
\end{eqnarray}
where $\epsilon_{F}:=(\hbar^{2}/2m)(6\pi^{2}N/V)^{2/3}$
is the  Fermi energy.
We can immediately see from eq. (\ref{p0}) that $P$ is non vanishing
even at absolute zero of temperature.

Although not that often introduced in the context of thermal physics,
the  Casimir effect is another example of counter-intuitive phenomena \cite{cas48},
which fits nicely in the framework of statistical mechanics 
\cite{lif56,meh67,bro69,bal78,sch78,amb83,mit00}.
The usual setup consists of two large perfect conductor neutral plates
facing each other at a distance $d$. 
As we will see in detail later in the text, as the absolute zero
of temperature $T$ is approached, 
and if contributions from outside the setup are negligible,
the perpendicular pressure between the plates behaves as
\begin{equation}
P=-\frac{\pi^{2}}{240}\frac{\hbar c}{d^{4}}+ \cdots,
\label{p1}
\end{equation}
where ellipses denotes exponential small corrections that 
drop to zero as $T\rightarrow 0$. Likewise the pressure in eq. (\ref{p0}),
the ``Casimir pressure'' in eq. (\ref{p1}) does not vanish at the absolute
zero of temperature. Still more surprising, the non vanishing eq. (\ref{p1}) 
is not a consequence of particles hitting the plates, but it is due to 
the very presence of the perfect conductor plates which cannot be ignored 
at this regime of temperature. Moreover, the negativeness of eq. (\ref{p1})
indicates that the plates are attracted to each other, and that seems to be  a very unusual behavior for the walls of a cavity containing blackbody radiation.

When temperature is taken to be arbitrarily large 
($T\rightarrow\infty$), within the same framework that leads to eq. (\ref{p1}),
now it results as leading contribution the ordinary blackbody radiation pressure, namely,
\begin{equation}
P=\frac{\pi^{2}}{45}\frac{(kT)^{4}}{(\hbar c)^{3}}+ \cdots,
\label{p2}
\end{equation}
where the main correction is a lower power term.
It should be noticed that eqs. (\ref{p1}) and (\ref{p2}) are two different
regimes of the same phenomenon. In fact,
there is a symmetry which allows one to go from high temperatures
to low temperatures expressions, and vice versa \cite{bro69,rav89}.

Clearly, the presence of $\hbar$ in eqs. (\ref{p0}) to (\ref{p2})
indicates the quantum
nature of these experimentally validated pressures
\footnote{At this point a word of caution is in order. 
In fact, perfect conductor plates are idealizations of real materials whose properties have eventually to be considered in experiments.
There is in the literature an issue of this nature regarding 
discrepancies in the subleading
contributions in eqs. (\ref{p1}) and (\ref{p2}) (see, e.g., 
refs. \cite{bre14} and \cite{bim17} for reviews).
Such an issue does not affect directly our results; but
it might do so if 
experimental validation were addressed.
A related matter is that certain models for real materials lead to violation of the
third law of thermodynamics (see \cite{kli22} and references therein) or even to negative entropy 
(see, e.g.,  refs. \cite{bor10,kim17}).
These features do not appear in the context of 
two perfect conducting parallel walls.}.
In fact, eq. (\ref{p2}) has played special role in the  birth of quantum mechanics as is very well known.

This paper is not concerned with how temperature modifies the Casimir effect.
At some extent,
it is quite the other way around. The goal here,
akin to the manner in refs. \cite{bro69} and \cite{ken80},
is to explore further 
how the Casimir effect modifies the ordinary thermodynamics of the blackbody  radiation. In order to fulfil this task, the work is organized as follows. 
In  Sec. \ref{tensor}, we present the asymptotic behaviors of the
ensemble average of the stress-energy-momentum tensor, 
$\left<T_\mu{}_\nu\right>$,
of the electromagnetic field in the background of two 
arbitrarily large and 
parallel perfect conductor plates, which are neutral.
The calculation of $\left<T_\mu{}_\nu\right>$ has been implemented in a
classic paper in the literature, ref. \cite{bro69}, where it is neatly shown
that the components of $\left<T_\mu{}_\nu\right>$ are consistent with   thermodynamics.

In Sec. \ref{quantities}, the energy density in $\left<T_\mu{}_\nu\right>$
is integrated between the plates to obtain the internal energy 
$U$. Then, using $U$ and the stress components in  $\left<T_\mu{}_\nu\right>$,
we obtain the expressions for the Helmholtz free energy $F$ at high temperatures (and/or large distances between the plates) and at low temperatures
 (and/or small distances between the plates). It turns out that 
 $F$ is a function of temperature $T$, area $A$ of the plates, and
distance $d$ between them. Accordingly, two different pressures arise:
$P_{\parallel}$ which is the pressure parallel to the plates, and
$P_{\perp}$ which is the pressure perpendicular to them. This anisotropy
is fully investigated at the two regimes of temperature mentioned above. 
Various quantities are explored and
compared with their counterparts in the ordinary blackbody thermodynamics.
It should be mentioned that our approach differs from that in previous accounts where only $P_{\perp}$ has been considered to study  blackbody thermodynamics in slabs (see, e.g., ref. \cite{mit00}).

Due to the presence of the Casimir effect, or perhaps more appropriately,
due to the presence of boundary conditions, certain thermodynamic quantities that are 
trivial or ill defined in ordinary blackbody thermodynamics become now non trivial or well defined.
In Secs.  \ref{quantities}  and  \ref{other quantities} 
we calculate such new quantities, exploring
their thermodynamic meanings. Thermodynamic stability is  addressed in Sec. \ref{other quantities}.

In Sec. \ref{processes}, we investigate classic thermodynamic processes
involving quantities obtained in the two previous sections, at both regimes of
temperature. Atypical behaviors are
spotted and confronted with the principles
of thermodynamics. It should be anticipated  that anisotropy plays a special 
role in this section.

We close the study with Sec. \ref{summary}, presenting a brief summary and
discussing a little further our results. Some new directions of investigation
are also considered in this final section.

\section{Brown-Maclay $\left<T_\mu{}_\nu\right>$}
\label{tensor}

More than five decades ago, Brown and Maclay revisited
the Casimir setup [in a Minkowski frame $(x^{0},x^{1},x^{2},x^{3})$, 
the plates are parallel to the plane $x^{1}x^{2}$,
one plate is sitting at $x^{3}=0$ and the other at $x^{3}=d$]
at finite temperature $T$ \cite{bro69}. Working with the Green's function, the authors obtained closed expressions for the asymptotic behaviors of the ensemble average of the  electromagnetic stress-energy-momentum tensor, $\left<T_\mu{}_\nu\right>$, whose nonvanishing
components are given as below.
\begin{eqnarray}
&&
\hspace{-2.0cm}
\frac{kTd}{\hbar c}\gg 1:
\nonumber
\\
&&
\langle T_{00}\rangle=\frac{\pi^{2}}{15}\frac{(kT)^{4}}{(\hbar c)^{3}},
\nonumber
\\
&&
\langle T_{11}\rangle=\langle T_{22}\rangle=\frac{\pi^{2}}{45}\frac{(kT)^{4}}{(\hbar c)^{3}}+\frac{\zeta(3)}{8\pi}\frac{kT}{d^{3}},
\nonumber
\\
&&
\langle T_{33}\rangle=\frac{\pi^{2}}{45}\frac{(kT)^{4}}{(\hbar c)^{3}}-\frac{\zeta(3)}{4\pi}\frac{kT}{d^{3}}.
\label{ht}
\end{eqnarray}

\begin{eqnarray}
&&
\hspace{-2.0cm}
\frac{kTd}{\hbar c}\ll 1:
\nonumber
\\
&&
\langle T_{00}\rangle=-\frac{\pi^2}{720}\frac{\hbar c}{d^{4}}+\frac{\zeta(3)}{\pi}\frac{(kT)^{3}}
{d\, (\hbar c)^{2}},
\nonumber
\\
&&
\langle T_{11}\rangle=\langle T_{22}\rangle=\frac{\pi^2}{720}\frac{\hbar c}{d^4}
+\frac{\zeta(3)}{2\pi}\frac{(kT)^{3}}{d\,(\hbar c)^{2}},
\nonumber
\\
&&
\langle T_{33}\rangle=-\frac{\pi^2}{240}\frac{\hbar c}{d^4}.
\label{lt}
\end{eqnarray}
Exponential small corrections have been neglected in both set of equations.

Although eqs. (\ref{ht}) and (\ref{lt}) correspond to the region between the plates
(which we call slab),
one can use these equations to learn about the behavior of $\left<T_\mu{}_\nu\right>$ everywhere.
When we set $d\rightarrow \infty$, eq. (\ref{ht}) tells us that there is only blackbody radiation
outside the slab. Now, keeping the temperature outside  
negligible compared with that inside, it follows that $\left<T_\mu{}_\nu\right>$ vanishes
outside the slab. Such an idealised picture will be used in the following sections.

\section{Thermodynamic quantities}
\label{quantities}
We shall consider the Casimir setup as a rectangular parallelepiped cavity 
with two squared parallel walls of area $A$, and the others of area $\sqrt{A}d$, where
\begin{equation}
d\ll\sqrt{A}.
\label{walls}
\end{equation}
Even though the calculations that led to eqs. (\ref{ht}) and (\ref{lt}) considered
only boundary conditions on the large walls, they can still apply to the perfect conducting
cavity, as long as $A$ is taken to be arbitrarily large \cite{lim07,gey08}.

Thermodynamics at the two regimes of temperature will be studied separately, starting with
high temperatures (and/or large $d$).
\subsection{$kTd/\hbar c\gg 1$}
\label{3ht}

By integrating the energy density 
$\langle T_{00}\rangle$ 
over space in eq. (\ref{ht}), we obtain the internal energy,
\begin{equation}
U=\frac{\pi^{2}}{15}\frac{(kT)^{4}}{(\hbar c)^{3}} Ad,
\label{htu}
\end{equation}
which one recognizes as the familiar blackbody radiation $U$ in a cavity of volume $Ad$
\cite{hua87,pat72,cal85,kel81}. Nevertheless, it should be recalled that eq. (\ref{htu}) hides exponential small corrections that have been neglected, as will be so in all formulae.

Considering their very definitions, the stress components of   
$\left<T_\mu{}_\nu\right>$ in eq. (\ref{ht}) are identified with the thermodynamic
pressures already mentioned in the text, 
\begin{equation}
P_{\parallel}=\frac{\pi^{2}}{45}\frac{(kT)^{4}}{(\hbar c)^{3}}+\frac{\zeta(3)}{8\pi}\frac{kT}{d^{3}}, \hspace{2.0cm} 
P_{\perp}=\frac{\pi^{2}}{45}\frac{(kT)^{4}}{(\hbar c)^{3}}-\frac{\zeta(3)}{4\pi}\frac{kT}{d^{3}}.
\label{htp}
\end{equation}
The Helmholtz free energy $F$ must satisfy,
 \begin{equation}
P_{\parallel}=-\frac{1}{d}\left(\frac{\partial F}{\partial A}\right)_{T,d}, 
\hspace{2.0cm} 
P_{\perp}=-\frac{1}{A}\left(\frac{\partial F}{\partial d}\right)_{T,A},
\label{p}
\end{equation}
and $U=\partial_{\beta}(\beta F)$, where $\beta:=1/kT$ \cite{cal85}.
Then, it follows from eqs. (\ref{htu}) to (\ref{p}) that, 
\begin{equation}
F=-\frac{\pi^{2}}{45}\frac{(kT)^{4}}{(\hbar c)^{3}} Ad
-\frac{\zeta(3)}{8\pi}\frac{kTA}{d^{2}},
\label{htf}
\end{equation}
agreeing with an early result where a different approach has been used \cite{bal78}.

One can easily check that the equation of state is given by,
\begin{equation}
\left(P_{\perp}+2P_{\parallel}\right)Ad=U,
\label{htes}
\end{equation}
becoming that of the ordinary hot radiation ($PV=U/3$) when
$d\rightarrow \infty$ for a given $T$.
It is worth noting that eq. (\ref{htes}) is simply another way of saying that 
$\left<T_\mu{}_\nu\right>$ is traceless.
In the context of Casimir physics \cite{mil01,bor01}, the negative term in the expression of $P_{\perp}$ 
[see eq. (\ref{htp})] is known as ``thermal Casimir effect'' \cite{sus11}. It diminishes the radiation pressure
on the large walls, whereas the corresponding correction in $P_{\parallel}$ increases the radiation
pressure on the other walls of the slab.

It follows from $S=-(\partial_{T} F)_{A,d}$ and eq. (\ref{htf}) that the entropy at
this regime is given by,
\begin{equation}
S=\frac{4\pi^{2}}{45}\left(\frac{kT}{\hbar c}\right)^{3}k Ad
+\frac{\zeta(3)}{8\pi}\frac{kA}{d^{2}}.
\label{hts}
\end{equation}
The correction to the blackbody entropy term in eq. (\ref{hts}) is sometime called ``Casimir's entropy'' \cite{rev97}.
It enhances $S$ and spoils its extensive character. There are claims stating that the corrections
in eq. (\ref{hts}) is ``classical'' since it does not contain $\hbar$.
However, we should note that the blackbody term, which is the leading contribution here, carries the factor $1/\hbar^{3}$ becoming arbitrarily large as 
$\hbar\rightarrow 0$. For future reference, below we recast $S$ as a function of 
$U$, $A$, and $d$, namely
\begin{equation}
S=\frac{4}{3}k
\left[\frac{\pi^{2}}{15 (\hbar c)^{3}}\right]^{1/4}\left(U^{3}Ad\right)^{1/4}
+\frac{\zeta(3)}{8\pi}\frac{kA}{d^{2}}.
\label{hte2}
\end{equation}
The entropy expressed in this way is known as fundamental relation \cite{cal85}, and it is a thermodynamic potential for which the following equalities are satisfied:
\begin{equation}
\frac{1}{T}=\left(\frac{\partial S}{\partial U}\right)_{A,d}, 
\hspace{1.0cm} 
P_{\parallel}=\frac{T}{d}\left(\frac{\partial S}{\partial A}\right)_{U,d},
\hspace{1.0cm} 
P_{\perp}=\frac{T}{A}\left(\frac{\partial S}{\partial d}\right)_{U,A}.
\label{epotential}
\end{equation}

Each pressure in eq. (\ref{htp}) is associated with an enthalpy, i.e.,
$H_{\parallel}=U+P_{\parallel}Ad$ and $H_{\perp}=U+P_{\perp}Ad$, whose expressions are given by
[see eq. (\ref{htu})],
\begin{equation}
H_{\parallel}=\frac{4\pi^{2}}{45}\frac{(kT)^{4}}{(\hbar c)^{3}}Ad+\frac{\zeta(3)}{8\pi}\frac{kTA}{d^{2}}, \hspace{2.0cm} 
H_{\perp}=\frac{4\pi^{2}}{45}\frac{(kT)^{4}}{(\hbar c)^{3}}Ad-\frac{\zeta(3)}{4\pi}\frac{kTA}{d^{2}}.
\label{hth}
\end{equation}
Furthermore, each pressure gives rise to  a Gibbs thermodynamic potential through 
$G_{\parallel}=F+P_{\parallel}Ad$ and $G_{\perp}=F+P_{\perp}Ad$. Noting eq. (\ref{htf}),
it follows that,
\begin{equation}
G_{\parallel}=0, \hspace{2.0cm} 
G_{\perp}=-\frac{3\zeta(3)}{8\pi}\frac{kTA}{d^{2}}.
\label{htg}
\end{equation}
It is rather surprising that $G_{\perp}$
does not vanish identically, in contrast with thermodynamics of ordinary blackbody radiation
\cite{kel81}. In order to fully appreciate this fact, we give $G_{\perp}$ as a function
of $T$, $A$ and $P_{\perp}$ as follows,
\begin{equation}
G_{\perp}=-\frac{3}{2}A
\left(\frac{\zeta(3)}{4\pi}kT\right)^{1/3}
\left[\frac{\pi^{2}}{45}\frac{(kT)^{4}}{(\hbar c)^{3}}-P_{\perp}\right]^{2/3}.
\label{htg2}
\end{equation}
One may check that the Gibbs potential in eq. (\ref{htg2}) 
consistently satisfies the following equations,
\begin{equation}
S=-\left(\frac{\partial G_{\perp}}{\partial T}\right)_{A,P_{\perp}}, 
\hspace{1.0cm} 
Ad=\left(\frac{\partial G_{\perp}}{\partial P_{\perp}}\right)_{T,A},
\hspace{1.0cm} 
P_{\perp}-P_{\parallel}=\frac{1}{d}
\left(\frac{\partial G_{\perp}}{\partial A}\right)_{T,P_{\perp}},
\label{gpotential}
\end{equation}
as it should. It is fair to say that nontrivial $G_{\perp}$ in eqs. (\ref{htg}) and (\ref{htg2})
arises due to the presence of Casimir's effect
\footnote{At this point one may be wondering why $G_{\parallel}$
in eq. (\ref{htg}) is still trivial. In fact $G_{\parallel}$ would be nontrivial if $A$ were not considered arbitrarily large 
(see, e.g., the toy models in ref. \cite{mor20}).}.

Another relevant thermodynamic quantity
is the heat capacity at constant volume
\footnote{
More correctly speaking, the heat capacity at constant $A$ and $d$
since we are dealing with a system with three thermodynamic parameters.}, i.e.,
$C_{V}=(\partial_{T} U)_{A,d}$. Noting eq. (\ref{htu}), we see that $C_{V}$, up to usual exponential small corrections, is the same as that
in ordinary blackbody thermodynamics, 
\begin{equation}
C_{V}=\frac{4\pi^{2}}{15}k\left(\frac{kT}{\hbar c}\right)^{3}Ad.
\label{htcv}
\end{equation}
However, the identity $C_{V}=3S$ is not any longer satisfied.
Instead,
\begin{equation}
C_{V}=3S-\frac{3\zeta(3)}{8\pi}\frac{kA}{d^{2}}
\label{htcvs}
\end{equation}
holds now. Heat capacity at constant pressure will be considered in Sec. \ref{other quantities}.

We turn now to the analyses of the regime of low temperatures (and/or small $d$).

\subsection{$kTd/\hbar c\ll 1$}
\label{3lt}

We will follow  the same steps of Sec. \ref{3ht}, but now using eq. (\ref{lt}).
The internal energy at temperatures near the absolute zero (and/or in narrow slabs)
is now given by, 
\begin{equation}
U=-\frac{\pi^2}{720}\frac{\hbar c}{d^{3}}A+\frac{\zeta(3)}{\pi}\frac{(kT)^{3}}
{(\hbar c)^{2}}A,
\label{ltu}
\end{equation}
where the negative (and dominant) term is the well known Casimir's energy \cite{mil01,bor01}.
The thermal correction in eq. (\ref{ltu}) has a peculiar interpretation.
It is precisely the ordinary blackbody expression for $U$ in $(2+1)$-dimensional spacetime
where the ``volume'' of the cavity is $A$ (see, e.g., ref. \cite{amb83}). At this regime, such a feature is shared by
many thermodynamic quantities.

It is instructive to compare the thermal energy in eq. (\ref{ltu})
with $U$ of a blackbody of volume $Ad$, at the same temperature $T$:
\begin{equation}
\frac{\zeta(3)}{\pi}\frac{(kT)^{3}}{(\hbar c)^{2}}A\,
\left(\frac{\pi^{2}}{15}\frac{(kT)^{4}}{(\hbar c)^{3}} Ad\right)^{-1}\approx
\frac{\hbar c}{kTd}\gg 1.
\label{3over4}
\end{equation}
Thus, eq. (\ref{3over4}) tells us that there is much more thermal energy in the 
Casimir slab than in a conventional blackbody slab. This remark is in qualitative agreement
\footnote{But only qualitative agreement. In ref. \cite{bim19} the thermal contribution in
eq. (\ref{ltu}) is almost linear on $T$ and lacks experimental evidence.} 
with
the study in ref. \cite{bim19}.

The stress components of   
$\left<T_\mu{}_\nu\right>$ in eq. (\ref{lt}) are the thermodynamic
pressures, 
\begin{equation}
P_{\parallel}=\frac{\pi^2}{720}\frac{\hbar c}{d^4}
+\frac{\zeta(3)}{2\pi d}\frac{(kT)^{3}}{(\hbar c)^{2}}, \hspace{2.0cm} 
P_{\perp}=-\frac{\pi^2}{240}\frac{\hbar c}{d^4},
\label{ltp}
\end{equation}
which also satisfy the equation of state eq. (\ref{htes}), with $U$ given by eq. (\ref{ltu}).
It is worth noting that $P_{\perp}$ in eq. (\ref{ltp})
is the attractive Casimir pressure in eq. (\ref{p1}). The dominant contribution in $P_{\parallel}$
is a repulsive Casimir pressure and its thermal correction is $1/d$ times the blackbody radiation pressure in $(2+1)$-dimensional background. The corresponding Helmholtz free energy $F$ and entropy $S$ are given by
\begin{equation}
F=-\frac{\pi^2}{720}\frac{\hbar c}{d^{3}}A-\frac{\zeta(3)}{2\pi}\frac{(kT)^{3}}
{(\hbar c)^{2}}A,
\hspace{2.0cm}
S=\frac{3\zeta(3)}{2\pi}\left(\frac{kT}{\hbar c}\right)^{2}kA,
\label{ltfs}
\end{equation}
agreeing once again with ref. \cite{bal78}.
Note that $S$ in eq. (\ref{ltfs}) is the entropy of
a $(2+1)$-dimensional blackbody.

It is rather puzzling that $S$ in eq. (\ref{ltfs}), that does not depend on $d$,
must lead to the Casimir pressure in eq. (\ref{p1}) [or eq. (\ref{ltp})] which does depend on $d$.
Nevertheless we should recall that in order to use $S$ as a thermodynamic potential to obtain pressure,
$S$ must be rewritten as a function of $U$, $A$ and $d$ [cf. eq. (\ref{hte2})], i.e.,
\begin{equation}
S=\frac{3}{2}k
\left[\frac{\zeta(3)}{\pi}\frac{A}{(\hbar c)^{2}}\right]^{1/3}
\left(U+\frac{\pi^{2}}{720}\frac{\hbar c}{d^{3}}A\right)^{2/3}.
\label{lts2}
\end{equation}
Then, by using eq. (\ref{epotential}),
we do end up with the Casimir pressure in eq. (\ref{ltp}).
In this sense eq. (\ref{lts2}) is a genuine ``Casimir's entropy''. For completeness, it shoud be
remarked that  $P_{\parallel}$ in eq. (\ref{ltp}) can also be reproduced by considering eq. (\ref{epotential}).

A last (speculative) remark regarding the entropy in eq. (\ref{ltfs}) is that, likewise the 
Bekenstein entropy of a black hole \cite{bek73}, it depends on the area $A$,
not on the volume of the slab. Such a loss of extensiveness has been spotted earlier in the text.

As in Sec. \ref{3ht}, corresponding to the pressures in eq. (\ref{ltp}), we have two enthalpies
\begin{equation}
H_{\parallel}=\frac{3\zeta(3)}{2\pi}\frac{(kT)^{3}}{(\hbar c)^{2}}A, \hspace{2.0cm} 
H_{\perp}=-\frac{\pi^{2}}{180}\frac{\hbar c}{d^{3}}A
+\frac{\zeta(3)}{\pi}\frac{(kT)^{3}}{(\hbar c)^{2}}A,
\label{lth}
\end{equation}
and two Gibbs potentials
\begin{equation}
G_{\parallel}=0, \hspace{2.0cm} 
G_{\perp}=-\frac{\pi^{2}}{180}\frac{\hbar c}{d^{3}}A
-\frac{\zeta(3)}{2\pi}\frac{(kT)^{3}}{(\hbar c)^{2}}A.
\label{ltg}
\end{equation}
We see that also at low temperatures (and/or in narrow slabs),
$G_{\perp}$ in eq. (\ref{ltg}) is nontrivial and carries information about the thermodynamic 
behavior of the system. 
Indeed, by expressing $G_{\perp}$
as a function of $T$, $A$ and $P_{\perp}$, i.e.,
\begin{equation}
G_{\perp}=-\frac{\pi^{2}}{180}\hbar c
\left(-\frac{240}{\pi^{2}}\frac{P_{\perp}}{\hbar c}\right)^{3/4}A
-\frac{\zeta(3)}{2\pi}\frac{(kT)^{3}}{(\hbar c)^{2}}A,
\label{ltg2}
\end{equation}
one can verify that eq. (\ref{gpotential}) still holds.
As already mentioned in the paper, this fact
differs from the ordinary blackbody thermodynamics where the Gibbs thermodynamic potential vanishes identically \cite{kel81}.

We close this section by making a couple of remarks.
The heat capacity at constant volume
for this regime [see eq. (\ref{ltu})] is given by
\begin{equation}
C_{V}=\frac{3\zeta(3)}{\pi}k\left(\frac{kT}{\hbar c}\right)^{2}A,
\label{ltcv}
\end{equation}
which is related to $S$ in eq. (\ref{ltfs}) by  $C_{V}=2S$ typical of a 
$(2+1)$-dimensional blackbody [cf. eqs. (\ref{htcv}) and (\ref{htcvs})].
By setting $T\rightarrow 0$ in eq. (\ref{ltfs}) it results that 
$S\rightarrow 0$ [and $C_{V}\rightarrow 0$], i.e., the third law of thermodynamics is satisfied,
as one expected.

\section{Other quantities}
\label{other quantities}

In ordinary blackbody thermodynamics, pressure $P$ depends only on 
temperature $T$ [see eq. (\ref{p2})]. Thus, any quantity defined as the rate of change of something with
respect to $T$ keeping $P$ constant, or the rate of change of something with respect to
$P$ keeping $T$ constant, is ill defined.
A well known example is the heat capacity at constant pressure
$C_{P}=(\partial_{T} H)_{P}$ \cite{pat72,kel81,rod89}.
As mentioned in ref. {\cite{rod89}}, some authors try to go around the 
problem by saying that $C_{P}=\infty$; but that leads quickly into difficulties. Now, we should recall that electromagnetic radiation
in a cavity is not just hot radiation, but it is hot radiation
in thermodynamic equilibrium with the walls of the cavity.

This section addresses analogs of ill defined blackbody  quantities that, in the present context, are
well defined due to the presence of  Casimir's effect. 

\subsection{$kTd/\hbar c\gg 1$}
\label{4ht}

By considering the first law of thermodynamics, i.e., 
$dQ=dU+P_{\parallel}a\, dA+P_{\perp}A\, da$,
it can be shown that the
analogs of $C_{P}$ are two heat capacities, 
$C_{P_{\parallel}}$ and $C_{P_{\perp}}$, given by
\begin{equation}
C_{P_{\parallel}}=\left(\frac{\partial H_{\parallel}}{\partial T}\right)_{P_{\parallel},d},
\hspace{2.0cm} 
C_{P_{\perp}}=\left(\frac{\partial H_{\perp}}{\partial T}\right)_{P_{\perp},A}.
\label{htcp}
\end{equation}
If $P_{\parallel}$ and $d$ are taken to be constant, eq. (\ref{htp}) tells
us that $T$ must also be constant, resulting that $C_{P_{\parallel}}$ in
eq. (\ref{htcp}) is ill defined 
\footnote{Again, the cause of this remaining ``pathology''
can be traced to the fact that $A$ is considered arbitrarily large 
(see, e.g., the toy models in ref. \cite{mor20}
where finite size one-dimensional boxes are considered).
Indeed, preliminary calculations in an investigation in progress show that $C_{P}$ is well defined in cubic and spherical cavities when the presence of the corresponding conducting walls are all taken into account.
} 
as it is $C_{P}$ in ordinary blackbody thermodynamics.
But that is not the case for $C_{P_{\perp}}$ in eq. (\ref{htcp}).
Looking at eq. (\ref{htp}) one can see that $P_{\perp}$
can be kept constant while $T$ and $d$ vary simultaneously with 
cancellation of their corresponding variations
\footnote{Note that this is only possible due to the presence of the
Casimir contribution in eq. (\ref{htp}).}.
We can use eq. (\ref{htp})
in eq. (\ref{hth}) to recast $H_{\perp}$ as a function of $T$, $P_{\perp}$ 
and $A$:
\begin{equation}
H_{\perp}=
A\left(\frac{\zeta(3)}{4\pi}kT\right)^{1/3}
\left[\frac{\pi^{2}}{15}\frac{(kT)^{4}}{(\hbar c)^{3}}+P_{\perp}\right]
\left[\frac{\pi^{2}}{45}\frac{(kT)^{4}}{(\hbar c)^{3}}-P_{\perp}\right]^{-1/3},
\label{hthp}
\end{equation}
following from eq. (\ref{htcp}) that
\begin{eqnarray}
C_{P_{\perp}}&=&-\frac{4\pi}{27\,\zeta(3)}k\frac{d^{2}}{A}
\left[\frac{C_{V}}{k}-\frac{3\zeta(3)}{4\pi}\frac{A}{d^{2}}\right]^{2}+C_{V}
\nonumber
\\
&=&-\frac{4\pi}{27\,\zeta(3)}\frac{d^{2}}{A}\frac{C_{V}^{2}}{k}
\left[1-\frac{495}{16\pi^{3}}\zeta(3)\left(\frac{\hbar c}{kTd}\right)^{3}+\cdots\right],
\label{htcpp}
\end{eqnarray}
where $C_{V}$ is given by eq. (\ref{htcv}). Thus
$C_{P_{\perp}}$ in eq. (\ref{htcpp}) is obviously finite and moreover it is negative (this negativeness will be addressed later). However, note that 
$|C_{P_{\perp}}/C_{V}|\approx (kTd/\hbar c)^{3}\gg 1$,
at this regime. We will show shortly that $C_{P_{\perp}}$
plays with other quantities the roles expected from a 
heat capacity at constant pressure \cite{cal85}.

In the slab, we can also define ``coefficients of thermal expansion'':
\begin{equation}
\alpha_{\parallel}:=\frac{1}{A}\left(\frac{\partial A}{\partial T}\right)
_{P_{\parallel},d},
\hspace{2.0cm}
\alpha_{\perp}:=\frac{1}{d}\left(\frac{\partial d}{\partial T}\right)
_{P_{\perp},A},
\label{hta}
\end{equation}
``isothermal compressibilities'':
\begin{equation}
(\kappa_{T})_{\parallel}:=-\frac{1}{A}\left(\frac{\partial A}{\partial P_{\parallel}}\right)_{T,d},
\hspace{2.0cm}
(\kappa_{T})_{\perp}:=-\frac{1}{d}\left(\frac{\partial d}{\partial P_{\perp}}\right)_{T,A},
\label{htkt}
\end{equation}
and ``adiabatic compressibilities'':
\begin{equation}
(\kappa_{S})_{\parallel}:=-\frac{1}{A}\left(\frac{\partial A}{\partial P_{\parallel}}\right)_{S,d},
\hspace{2.0cm}
(\kappa_{S})_{\perp}:=-\frac{1}{d}\left(\frac{\partial d}{\partial P_{\perp}}\right)_{S,A}.
\label{htks}
\end{equation}
The definitions in eqs. (\ref{hta}) and (\ref{htkt}) are analogs of 
ill defined quantities in the familiar blackbody thermodynamics,
whereas eq. (\ref{htks}) are corrections to the well defined blackbody
adiabatic compressibility, as we will see below.

Examining eqs. (\ref{hta}) and (\ref{htkt}) we can check that, 
likewise  $C_{P_{\parallel}}$ in eq. (\ref{htcp}),
$\alpha_{\parallel}$ and $(\kappa_{T})_{\parallel}$
are both ill defined. Nevertheless, 
$\alpha_{\perp}$ and $(\kappa_{T})_{\perp}$ are well defined, and by
considering eq. (\ref{htp}) one is led to
\begin{equation}
\alpha_{\perp}=\frac{H_{\perp}}{2TG_{\perp}},
\hspace{2.0cm}
(\kappa_{T})_{\perp}=\frac{Ad}{2G_{\perp}},
\label{htak}
\end{equation}
with the enthalpy $H_{\perp}$ and the Gibbs potential $G_{\perp}$
given by eqs. (\ref{hth}) and (\ref{htg}). Indeed eq. (\ref{htak})
carries information about the thermodynamic behavior of the system
at hight temperatures (and/or large $d$).
For example, since $\alpha_{\perp}$ is negative,
it results that the slab becomes thinner as it warms when $P_{\perp}$
and $A$ are kept constant 
[cf. its definition in eq (\ref{hta})]. Physical interpretation of 
$(\kappa_{T})_{\perp}$ will be addressed later on in the text.
For a check of consistency, we may appreciate that the 
following identity involving the difference of heat capacities is satisfied
\cite{hua87,cal85},
namely:
\begin{equation}
C_{P_{\perp}}-C_{V}=\frac{TAd\, \alpha_{\perp}^{2}}{(\kappa_{T})_{\perp}}.
\label{cp-cv}
\end{equation}

In order to calculate the quantities in eq. (\ref{htks}), we take
$S$ constant in eq. (\ref{hts}) and use it in the expressions for the pressures
in eq. (\ref{htp}), yielding

\begin{eqnarray}
(\kappa_{S})_{\parallel}&=&\frac{135}{4\pi^{2}}\frac{(\hbar c)^{3}}{(kT)^{4}}
\left[1+\frac{45\,\zeta(3)}{32\pi^{3}}\left(\frac{\hbar c}{kTd}\right)^{3}\right]^{-2},
\label{htkspa}
\\
(\kappa_{S})_{\perp}&=&\frac{135}{4\pi^{2}}\frac{(\hbar c)^{3}}{(kT)^{4}}
\left[\left(1-\frac{45\,\zeta(3)}{16\pi^{3}}\left(\frac{\hbar c}{kTd}\right)^{3}\right)^{2}-\frac{405\,\zeta(3)}{16\pi^{3}}\left(\frac{\hbar c}{kTd}\right)^{3}\right]^{-1},
\label{htkspp}
\end{eqnarray}
where one recognizes the factor out of the right brackets in eqs. (\ref{htkspa}) and (\ref{htkspp}) as $\kappa_{S}=3/4P$ of an ordinary blackbody.
We have checked that
\begin{equation}
\frac{(\kappa_{S})_{\perp}}{(\kappa_{T})_{\perp}}=
\frac{C_{V}}{C_{P_{\perp}}}
\label{kspe/ktpe}
\end{equation}
holds as expected from thermodynamics \cite{hua87,cal85}. Thus we might wonder if 
\begin{equation}
\frac{(\kappa_{S})_{\parallel}}{(\kappa_{T})_{\parallel}}=
\frac{C_{V}}{C_{P_{\parallel}}}
\label{kspa/ktpa}
\end{equation}
is also satisfied. Whereas $(\kappa_{S})_{\parallel}$ and $C_{V}$ 
in eq. (\ref{kspa/ktpa}) are well defined quantities in the slab; $(\kappa_{T})_{\parallel}$ and
$C_{P_{\parallel}}$ are not as it has been already mentioned.
However, eq. (\ref{kspa/ktpa}) should indeed be satisfied in a slab where the area $A$
were not taken arbitrarily large.

Noting eq. (\ref{htcv}) one has that $C_{V}$ is positive, which is the criterion for
thermal stability of a thermodynamic system. Nevertheless, eq. (\ref{htak}) tells us
that $(\kappa_{T})_{\perp}<0$ [or we note that $C_{P_{\perp}}<0$, cf. eqs. (\ref{htcpp}) 
and (\ref{cp-cv})], 
therefore violating 
the criterion for mechanical stability \cite{cal85}.
It turns out that, at the regime of high temperatures (and/or large $d$),
electromagnetic radiation in the slab is not thermodynamic stable.
This conclusion seems to be in conflict with ref. \cite{kel81} where it is claimed that
blackbody radiation in a cavity is stable. We would rather say that it depends
very much on the ``geometry'' of the cavity, and on the nature of its walls.

At this point it is worth remarking that, as put in ref. \cite{hua87}, the quantities in eqs. (\ref{hta}) to (\ref{htks})
are particular important as they are experimentally measurable.

\subsection{$kTd/\hbar c\ll 1$}
\label{4lt}

Looking at the expressions found in Sec. \ref{3lt},
we wish now to calculate the well defined quantities $C_{P_{\perp}}$,
$\alpha_{\perp}$, $(\kappa_{T})_{\perp}$, $(\kappa_{S})_{\parallel}$ and
$(\kappa_{S})_{\perp}$ (as in Sec. \ref{4ht}, the remaining quantities are ill defined),
when the temperature is low (and/or the slab is narrow).

Corresponding to eq. (\ref{hthp}), one has that
\begin{equation}
H_{\perp}=-\frac{\pi^{2}}{180}\hbar c
\left(-\frac{240}{\pi^{2}}\frac{P_{\perp}}{\hbar c}\right)^{3/4}A
+\frac{\zeta(3)}{\pi}\frac{(kT)^{3}}{(\hbar c)^{2}}A,
\label{lthp}
\end{equation}
where eqs. (\ref{ltp}) and (\ref{lth}) have been used. It follows then from eqs. (\ref{htcp})
and (\ref{lthp}) the equality
\begin{equation}
C_{P_{\perp}}=C_{V},
\label{cp=cv}
\end{equation}
with $C_{V}$ given in eq. (\ref{ltcv})
\footnote{Likewise $C_{V}$,
note that $C_{P_{\perp}}$ also 
vanishes as the absolute zero of temperature is approached.}. 
Then one immediately sees from eq. (\ref{cp-cv}) that
\begin{equation}
\alpha_{\perp}=0,
\label{vanishing-a}
\end{equation}
and from eq. (\ref{kspe/ktpe}) that
\begin{equation}
(\kappa_{S})_{\perp}=(\kappa_{T})_{\perp}=\frac{1}{4P_{\perp}},
\label{kspe=ktpe}
\end{equation}
where the last equality in eq. (\ref{kspe=ktpe}) has been obtained by considering
eqs. (\ref{ltp}) and (\ref{htkt}). It should be noticed that we could have anticipated
eq. (\ref{vanishing-a}) by looking
at eq. (\ref{hta}): as $d$ is a function of $P_{\perp}$ only [cf. eq. (\ref{ltp})],
the coefficient of thermal expansion $\alpha_{\perp}$ must vanish identically.
Another remark before going to the next section. We see 
again that, as $C_{V}>0$ but $(\kappa_{T})_{\perp}<0$, at low temperatures (and/or in narrow slabs) the system is thermally stable; but mechanically unstable.

\section{Thermodynamic processes}
\label{processes}
At this section some features of certain thermodynamic processes will be considered where  $d$ and $A$ are allowed to vary separately with one of them kept fixed. Such an approach will, hopefully, make the discussion pedagogical.

\subsection{Processes with $A$ fixed}
\label{Afixed}

\subsubsection{$kTd/\hbar c\gg 1$}
\label{51ht}

Let us begin with isothermal processes at the regime of high temperatures (and/or in wide slabs).

\vspace{0.3cm}

{\it Constant $T$}

\vspace{0.3cm}

The system goes reversibly from $d=d_{i}$ to $d=d_{f}$.
The work done by the system is given by integrating $P_{\perp}A$, noting eq. (\ref{htp}),
from $d=d_{i}$ to $d=d_{f}$, yielding

\begin{equation}
W=\frac{\pi^{2}}{45}\frac{(kT)^{4}}{(\hbar c)^{3}}A\left(d_{f}-d_{i}\right)
+A\frac{\zeta(3)}{8\pi}kT\left(\frac{1}{d_{f}^{2}}-\frac{1}{d_{i}^{2}}\right).
\label{ahtwt}
\end{equation}
Correspondingly, the heat $Q=\int_{d_{i}}^{d_{f}}T\, dS$
absorbed by the system is [see eq. (\ref{hts})]

\begin{equation}
Q=\frac{4\pi^{2}}{45}\frac{(kT)^{4}}{(\hbar c)^{3}}A\left(d_{f}-d_{i}\right)
+A\frac{\zeta(3)}{8\pi}kT\left(\frac{1}{d_{f}^{2}}-\frac{1}{d_{i}^{2}}\right).
\label{ahtqt}
\end{equation}
Note that in an expansion ($d_{i}<d_{f}$) the corrections to the positive blackbody contributions in eqs. (\ref{ahtwt}) and (\ref{ahtqt}) are negative.
Note also that
\begin{equation}
Q-W=\Delta U,
\label{firstlaw}
\end{equation} 
as should be [cf. eq. (\ref{htu})], and that roughly $3Q/4$ goes  to
$U$ to keep the temperature constant whereas only $Q/4$ is spent in work
doing by the system.

A well known fact of thermodynamics involving the enthalpy $H$ and $Q$ is that, in a process where pressure $P$ is kept constant, then
\begin{equation}
\Delta H= Q.
\label{dh=q}
\end{equation}
In ordinary blackbody thermodynamics, constant $T$ implies constant $P$, and therefore
eq. (\ref{dh=q}) holds \cite{kel81}. However, noting eq. (\ref{hth}), we have for the
process under consideration that [cf. eq. (\ref{ahtqt})]
\begin{equation}
\Delta H_{\perp}=\frac{4\pi^{2}}{45}\frac{(kT)^{4}}{(\hbar c)^{3}}A\left(d_{f}-d_{i}\right)
-A\frac{\zeta(3)}{4\pi}kT\left(\frac{1}{d_{f}^{2}}-\frac{1}{d_{i}^{2}}\right)
\neq Q.
\label{dhneq}
\end{equation}
The reason for the outcome in eq. (\ref{dhneq}) can be easily traced to the presence
of Casimir's contribution in eq. (\ref{htp}): 
constant temperature does not necessarily imply constant pressure.
For completeness we mention that 
\begin{equation}
\Delta F=-W
\label{df=-w}
\end{equation}
holds as can be promptly verified.

\vspace{0.3cm}

{\it Constant $U$}

\vspace{0.3cm}

In a reversible process with $U=U_{0}$ throughout
[note that $U$ in eq. (\ref{htu}) is positive], eqs. (\ref{ahtwt}) and (\ref{ahtqt})
give place to
\begin{equation}
W=Q=\frac{U_{0}}{3}\ln\left(\frac{d_{f}}{d_{i}}\right)+
\frac{\zeta(3)}{9\pi^{3/2}}\left[15\, A^{3}(\hbar c)^{3}U_{0}\right]^{1/4}
\left(\frac{1}{d_{f}^{9/4}}-\frac{1}{d_{i}^{9/4}}\right),
\label{ahtwu}
\end{equation}
where the term containing $\zeta(3)$ is the subleading contribution.
One sees from eq. ({\ref{ahtwu}}) that $Q>0$ implies $d_{i}<d_{f}$, i.e.,
it results an expansion as expected from any gas. Nevertheless, we will see shortly that
guesses are not always fulfilled.

Assume the following process where $U$ is kept constant, say $U=U_{0}$. 
The walls of area $A$,
initially with $d=d_{i}$, are allowed to move freely for a short while when they are
held fixed again with $d=d_{f}$. Since the leading contribution in $P_{\perp}$ is positive
[cf. eq. (\ref{htp})], 
the system undergoes a spontaneous expansion
\footnote{By taking into account conservation of linear momentum of the electromagnetic field and its sources, one can show that positive and negative pressures correspond, respectively,  to expansion and contraction \cite{jac75}.}
, i.e., $d_{i}<d_{f}$.
Consider that before the expansion $T=T_{i}$, and that after $T=T_{f}$. 
Now, it can be easily shown that
\footnote{As we are interested in state functions, an irreversible process can be  replaced by a reversible one and  then partial derivatives carry useful  information.}
[see eq. (\ref{htu})]
\begin{equation}
\left(\frac{\partial T}{\partial d}\right)_{U,A}=-\frac{T}{4d}<0,
\label{dt/dd}
\end{equation}
and therefore the system cools down, i.e., $T_{i}>T_{f}$,
resembling the free expansion of 
an ordinary photon gas
\cite{kel81}. Of course entropy
increases: noting eq. (\ref{epotential}), recall that $P_{\perp}>0$ and that $d$ increases. 
By using eq. (\ref{dt/dd}) in eq. (\ref{htp}), one can show that both
pressures diminish, namely
\begin{eqnarray}
\left(\frac{\partial P_{\parallel}}{\partial d}\right)_{U,A}&=&
-\frac{U_{0}}{3Ad^{2}}\left[1+\frac{585\, \zeta(3)}{32\pi^{3}}\left(\frac{\hbar c}{kTd}\right)^{3}\right]<0,
\label{htpa}
\\
\left(\frac{\partial P_{\perp}}{\partial d}\right)_{U,A}&=&
-\frac{U_{0}}{3Ad^{2}}\left[1-\frac{585\, \zeta(3)}{16\pi^{3}}\left(\frac{\hbar c}{kTd}\right)^{3}\right]<0.
\label{htpe}
\end{eqnarray}

\vspace{0.3cm}

{\it Constant $S$}

\vspace{0.3cm}

Let us examine a reversible adiabatic process in which the system goes from
$d=d_{i}$ to $d=d_{f}$. We note that $Q=0$,
\begin{equation}
\Delta U= -W,
\label{du=-w}
\end{equation}
and $S$ is kept constant. We can determine $W$ simply by using eq. (\ref{du=-w})
with eqs. (\ref{htu}) and (\ref{htp}), yielding
\begin{equation}
W=3A\left[(P_{\perp})_{i}\, d_{i}-(P_{\perp})_{f}\, d_{f}
+\frac{\zeta(3)}{4\pi}k\left(\frac{T_{i}}{d_{i}^{2}}-\frac{T_{f}}{d_{f}^{2}}\right)
\right],
\label{ahtws}
\end{equation}
where the subscripts $i$ and $f$ refer to the initial and final states, respectively.
By ignoring the term that carries $\zeta(3)$
in eq. (\ref{ahtws})
we are left with the familiar adiabatic $W$ 
in ordinary blackbody thermodynamics \cite{kel81}.

It is worth remarking that the quantities $(Ad)T^{3}$ and $P_{\perp}(Ad)^{4/3}$ are not constant, but approximately constant only
[see, e.g., eq. (\ref{hts})], due to the presence of Casimir's effect.

\vspace{0.3cm}

{\it Constant $H_{\perp}$}

\vspace{0.3cm}

We are interested in calculating the Joule-Thomson coefficient
\begin{equation}
\mu_{\perp}:=
\left(\frac{\partial T}{\partial P_{\perp}}\right)_{H_{\perp},A},
\label{joule-thomsona1}
\end{equation}
which may have experimental relevance \cite{cal85}.
By considering $H_{\perp}$ in eq. (\ref{hth}) constant and using eqs. (\ref{htp}) and 
(\ref{joule-thomsona1}), we can show that
\begin{eqnarray}
\mu_{\perp}&=&\frac{Ad}{C_{P_{\perp}}}
\left(T\alpha_{\perp}-1\right)
\label{joule-thomson-identity}
\\
&=&
\frac{45}{4\pi^{2}k}\left(\frac{\hbar c}{kT}\right)^{3}
\left[1+\frac{585\, \zeta(3)}{16\pi^{3}}\left(\frac{\hbar c}{kTd}\right)^{3}
+ \cdots\right]>0,
\label{joule-thomsona2}
\end{eqnarray}
where $C_{P_{\perp}}$ and $\alpha_{\perp}$ are given in eqs. (\ref{htcpp})
and (\ref{htak}). It should be pointed out that the factor out of the right brackets
in eq. (\ref{joule-thomsona2}) is precisely the known Joule-Thomson coefficient
of a photon gas \cite{kel81}. And, alike the photon gas, as 
eq. (\ref{joule-thomsona2}) shows,
the system cools down when the pressure drops.

\subsubsection{$kTd/\hbar c\ll 1$}
\label{51lt}

Now we turn to processes at the regime of low temperatures (and/or in narrow slabs).

\vspace{0.3cm}

{\it Constant $T$ and $S$}

\vspace{0.3cm}

Noticing eqs. (\ref{ltp}) and (\ref{ltfs}), 
corresponding to eqs. (\ref{ahtwt}) and (\ref{ahtqt}) 
now we have that
\begin{equation}
W=\frac{\pi^{2}}{720}\hbar c\, A\left(\frac{1}{d_{f}^{3}}-\frac{1}{d_{i}^{3}}\right),
\hspace{2.0cm} 
Q=0,
\label{altwqt}
\end{equation}
and, as expected, eq. (\ref{du=-w}) is satisfied [cf. eq. (\ref{ltu})]. 
Examining eq. (\ref{altwqt}) one sees that in an expansion ($d_{i}<d_{f}$)
work is done on the system increasing $U$, and that the process is also adiabatic \footnote{In fact nearly adiabatic. This is a consequence of the neglecting of the already mentioned exponential small corrections.}.
Thus, when $A$ is fixed, heat is not required to keep the temperature constant differently to what happens
at the other regime just discussed. Once more, eq. (\ref{dh=q}) is not satisfied
for $H_{\perp}$
[see eq. (\ref{lth})] and eq. (\ref{df=-w}) still holds [see eq. (\ref{ltfs})].

We remark some features proper of constant $S$ [cf. eq. (\ref{ltfs})].
Noting the pressures in eq. (\ref{ltp}), it follows that
\begin{equation}
\left(P_{\parallel}+\frac{P_{\perp}}{3}\right)A^{3/2}d={\rm const.},
\label{ltadiabatic}
\end{equation}
which is the analog of 
\begin{equation}
PV^{4/3}={\rm const.}
\label{adiabatic}
\end{equation}
in the  usual blackbody thermodynamics
\cite{pat72}. 
However, while in eq. (\ref{adiabatic}) $PV$ always diminishes as
$V$ increases; eq. (\ref{ltadiabatic}) contemplates other possibilities.
Perhaps the best way of seeing that is to recast it as
\begin{equation}
\left(P_{\parallel}-\frac{|P_{\perp}|}{3}\right)A^{3/2}d={\rm const.},
\label{ltadiabatic2}
\end{equation}
highlighting the fact that $P_{\perp}<0$ [see eq. (\ref{ltp})].
We will come back to eqs. (\ref{ltadiabatic}) and (\ref{ltadiabatic2}) shortly \footnote{It should be remarked that 
eqs. (\ref{ltadiabatic}) and (\ref{ltadiabatic2}) hold even when $A$
 and/or $d$ are not fixed.}.

One can also check that  [cf. eq. (\ref{ahtws})]
\begin{equation}
W=\frac{A}{3}\left[(P_{\perp})_{i}\, d_{i}-(P_{\perp})_{f}\, d_{f}
\right].
\label{altws}
\end{equation}

\vspace{0.3cm}

{\it Constant $U$}

\vspace{0.3cm}

Noting eq. (\ref{ltp}), when we integrate $P_{\perp}A$ for fixed $A$
from $d=d_{i}$ to $d=d_{f}$, it results again $W$ in eq. (\ref{altwqt}), 
but now since $U$ is constant, $W=Q$. Therefore if $Q>0$, $W>0$
and consequently $d_{i}>d_{f}$. That is, the system contracts when it is heated up
while keeping $U$ constant, pulling parts  of the environment to which the system
may be attached.
This fact is in sharp contrast with the behavior of a typical gas
[recall the discussion following eq. (\ref{ahtwu})].
It is more typical of a rubber band \cite{cal85}.

Let us consider again a spontaneous process from $d_{i}$ to $d_{f}$ where
$U$ in eq. (\ref{ltu}) is kept constant. As $P_{\perp}<0$ [cf. eq. (\ref{ltp})],
the system undergoes a contraction, i.e., $d_{i}>d_{f}$.
Looking at eq. (\ref{ltu}), it follows that
\begin{equation}
\left(\frac{\partial T}{\partial d}\right)_{U,A}=
-\frac{\pi^{3}}{720\,\zeta(3)}\,\frac{(\hbar c)^{3}}{kd^{2}(kTd)^{2}}<0,
\label{2dt/dd}
\end{equation}
and thus $T_{i}<T_{f}$. In other words the system warms up in this spontaneous contraction.
Of course $S$ increases [see, e.g., eq. (\ref{ltfs})].

One can use eq. (\ref{2dt/dd}) in eq. (\ref{ltp}), to check that 
[cf. eqs. (\ref{htpa}) and (\ref{htpe})]
\begin{equation}
\left(\frac{\partial P_{\parallel}}{\partial d}\right)_{U,A}=
-\frac{11\,\pi^2}{1440}\frac{\hbar c}{d^5}
-\frac{\zeta(3)}{2\pi d^{2}}\frac{(kT)^{3}}{(\hbar c)^{2}}<0, \hspace{2.0cm} 
\left(\frac{\partial P_{\perp}}{\partial d}\right)_{U,A}=
\frac{\pi^2}{60}\frac{\hbar c}{d^5}>0.
\label{dp/dd}
\end{equation}
The inequalities in eq. (\ref{dp/dd}) show that whereas $P_{\parallel}$
grows, $P_{\perp}$ diminishes in the free contraction.

\vspace{0.3cm}

{\it Constant $H_{\perp}$}

\vspace{0.3cm}

In order to determine the Joule-Thomson coefficient 
we can use the identity in eq. (\ref{joule-thomson-identity})
noting eqs. (\ref{cp=cv}) and (\ref{vanishing-a}), yielding
\begin{equation}
\mu_{\perp}=
-\frac{\pi}{3\,\zeta(3)}\frac{d}{k}\left(\frac{\hbar c}{kT}\right)^{2}<0.
\label{joule-thomsona3}
\end{equation}
That is, the system warms up when the pressure drops.
Now, noticing eqs. (\ref{joule-thomsona2}) and (\ref{joule-thomsona3})
one sees that $\mu_{\perp}$ change sign in passing from one regime to the other
indicating that there is an inversion temperature along the way \cite{cal85}.

\subsection{Processes with $d$ fixed}
\label{dfixed}

\subsubsection{$kTd/\hbar c\gg 1$}
\label{52ht}

As we did above let us begin 
with isothermal processes at the regime of high temperatures (and/or in wide slabs).

\vspace{0.3cm}

{\it Constant $T$}

\vspace{0.3cm}

The system goes reversibly from $A=A_{i}$ to $A=A_{f}$, keeping $d$ fixed now.
Noting $P_{\parallel}$  in   eq. (\ref{htp}), integration over $A$ yields
\begin{equation}
W=\frac{\pi^{2}}{45}\frac{(kT)^{4}}{(\hbar c)^{3}}\left(A_{f}-A_{i}\right)d
+\frac{\zeta(3)}{8\pi}\frac{kT}{d^{2}}\left(A_{f}-A_{i}\right).
\label{dhtwt}
\end{equation}
The corresponding $Q$ is obtained as usual by integrating $T$ over $S$ given in eq. (\ref{hts}),
i.e.,
\begin{equation}
Q=\frac{4\pi^{2}}{45}\frac{(kT)^{4}}{(\hbar c)^{3}}\left(A_{f}-A_{i}\right)d
+\frac{\zeta(3)}{8\pi}\frac{kT}{d^{2}}\left(A_{f}-A_{i}\right),
\label{dhtqt}
\end{equation}
and thus eq. (\ref{firstlaw}) is satisfied as expected [cf. eq. (\ref{htu})].
It should be remarked that, unlike their counterparts in eqs. (\ref{ahtwt}) and (\ref{ahtqt}), the corrections to the ordinary blackbody terms in eqs.
(\ref{dhtwt}) and (\ref{dhtqt}) are positive  in expansions (i.e., when $A_{i}<A_{f}$).

By examining $P_{\parallel}$  in   eq. (\ref{htp}), one sees that, as $d$ is kept fixed
here,
the presence of Casimir's contribution does not spoil the fact that constant $T$
implies constant $P_{\parallel}$. Therefore eq. (\ref{dh=q}) holds for $H_{\parallel}$
[cf. eq. (\ref{dhneq})]. It is a simple task to show that eq. (\ref{df=-w}) also holds.

\vspace{0.3cm}

{\it Constant $U$}

\vspace{0.3cm}

In a reversible process with $U=U_{0}$,
corresponding to eq. (\ref{ahtwu}) we find that
\begin{equation}
W=Q=\frac{U_{0}}{3}\ln\left(\frac{A_{f}}{A_{i}}\right)+
\frac{\zeta(3)}{6\pi^{3/2}}\left[15 \left(\frac{\hbar c}{d^{3}}\right)^{3}U_{0}\right]^{1/4}
\left(A_{f}^{3/4}-A_{i}^{3/4}\right),
\label{dhtwu}
\end{equation}
where the first term in the second equality is the known blackbody expression
for $W$ \cite{kel81}, followed by a Casimir's correction. 
Again, $Q>0$ leads to an expansion, i.e., $A_{i}<A_{f}$, with no surprises.

For the spontaneous process from $A_{i}$ to $A_{f}$, considering that 
$P_{\parallel}$ in eq. (\ref{htp}) is positive,
the system undergoes an expansion.
By using eq. (\ref{htu}) it follows that [cf. eq (\ref{dt/dd})]
\begin{equation}
\left(\frac{\partial T}{\partial A}\right)_{U,d}=-\frac{T}{4A}<0,
\label{dt/dA}
\end{equation}
and we therefore have a cooling process. Looking at, e.g., the expression for
$S$ in eq. (\ref{hte2}), one sees that $S$ increases as it should.
By using eq. (\ref{dt/dA}) in eq. (\ref{htp}), it follows that
\begin{eqnarray}
\left(\frac{\partial P_{\parallel}}{\partial A}\right)_{U,d}&=&
-\frac{U_{0}}{3A^{2}d}\left[1+\frac{45\, \zeta(3)}{32\pi^{3}}\left(\frac{\hbar c}{kTd}\right)^{3}\right]<0,
\label{dhtpa}
\\
\left(\frac{\partial P_{\perp}}{\partial A}\right)_{U,d}&=&
-\frac{U_{0}}{3A^{2}d}\left[1-\frac{45\, \zeta(3)}{16\pi^{3}}\left(\frac{\hbar c}{kTd}\right)^{3}\right]<0.
\label{dhtpe}
\end{eqnarray}
That is, eqs. (\ref{dhtpa}) and (\ref{dhtpe}) tell us that both
pressures decrease 
[cf. eqs. (\ref{htpa}) and (\ref{htpe})].

\vspace{0.3cm}

{\it Constant $S$}

\vspace{0.3cm}

Let us consider a reversible adiabatic process in which the system goes from
$A=A_{i}$ to $A=A_{f}$. Once more $Q=0$, eq. (\ref{du=-w}) still holds
and $S$ is kept constant. Then, as before, $W$ follows from eq. (\ref{du=-w})
with eqs. (\ref{htu}) and (\ref{htp}), i.e. [cf. eq. (\ref{ahtws})],
\begin{equation}
W=3d\left[(P_{\parallel})_{i}\, A_{i}-(P_{\parallel})_{f}\, A_{f}
-\frac{\zeta(3)}{8\pi}
\frac{k}{d^{3}}
\left(A_{i}T_{i}-A_{f}T_{f}\right)
\right].
\label{dhtws}
\end{equation}
Again, the familiar expression in blackbody thermodynamics arises by dropping
the term with $\zeta(3)$ in eq. (\ref{dhtws}) \cite{kel81}.
We should mention that the comments on the paragraph next to that 
of  eq. (\ref{ahtws}) still apply.

\vspace{0.3cm}

{\it Constant $H_{\parallel}$}

\vspace{0.3cm}

The appropriate Joule-Thomson coefficient in the present case is
\begin{equation}
\mu_{\parallel}:=
\left(\frac{\partial T}{\partial P_{\parallel}}\right)_{H_{\parallel},d}.
\label{joule-thomsond1}
\end{equation}
By taking $H_{\parallel}$ in eq. (\ref{hth}) constant and using eqs. (\ref{htp}) and 
(\ref{joule-thomsond1}), it follows that
\begin{eqnarray}
\mu_{\parallel}&=&
\frac{45}{4\pi^{2}k}\left(\frac{\hbar c}{kT}\right)^{3}
\left[1+\frac{45\, \zeta(3)}{32\pi^{3}}\left(\frac{\hbar c}{kTd}\right)^{3}\right]^{-1}
\label{joule-thomsond3}
\\
&=&
\frac{45}{4\pi^{2}k}\left(\frac{\hbar c}{kT}\right)^{3}
\left[1-\frac{45\, \zeta(3)}{32\pi^{3}}\left(\frac{\hbar c}{kTd}\right)^{3}
+ \cdots\right]>0,
\label{joule-thomsond2}
\end{eqnarray}
which should be compared with eq. (\ref{joule-thomsona2}) for $\mu_{\perp}$.
An interesting point arises here. Whereas $\mu_{\perp}$ can be expressed
as in eq. (\ref{joule-thomson-identity}), such an identity
is not available for $\mu_{\parallel}$ simply because $C_{P_{\parallel}}$ and $\alpha_{\parallel}$ are ill defined. As we have pointed out previously, this is a ``defect'' caused by considering $A$ arbitrarily large.
It would also affect $\mu_{\perp}$ if $d$ were taken arbitrarily large too, i.e.,
if only the familiar blackbody expressions had been taken into account.
Perhaps an exercise would clarify this issue further. By considering
$d$ arbitrarily large, the terms carrying $\zeta(3)$ in eq. (\ref{ht})
would be negligible. That would correspond to replace $\zeta(3)$ in eq. (\ref{htcpp}) by zero giving an ill defined $C_{P_{\perp}}$.

Note that the comments closing the paragraph of eq. (\ref{joule-thomsona2}) also apply to $\mu_{\parallel}$.

\subsubsection{$kTd/\hbar c\ll 1$}
\label{52lt}

Turning to processes at the regime of low temperatures (and/or in narrow slabs), it should be noticed from the beginning [see eq. (\ref{ltfs})] that
constant $T$ does not imply necessarily  constant $S$, since now $A$ is not fixed.
\vspace{0.3cm}

{\it Constant $T$}

\vspace{0.3cm}

Once more, by using  eqs. (\ref{ltp}) and (\ref{ltfs}) we are led to
the following expressions for the work done by the system,
\begin{equation}
W=\left[\frac{\pi^{2}}{720}\frac{\hbar c}{d^{3}}+\frac{\zeta(3)}{2\pi}
\frac{(kT)^{3}}{(\hbar c)^{2}}\right]\left(A_{f}-A_{i}\right),
\label{dltwt}
\end{equation}
and the heat absorbed by the system,
\begin{equation}
Q=\frac{3\zeta(3)}{2\pi}
\frac{(kT)^{3}}{(\hbar c)^{2}}\left(A_{f}-A_{i}\right).
\label{dltqt}
\end{equation}
Thus, thinking about an expansion ($A_{i}<A_{f}$),
eqs. (\ref{dltwt}) and (\ref{dltqt}) say that $W\gg Q$,
i.e., the work done by the system is essentially at the expenses of 
$U$  [see eqs. (\ref{ltu}) and (\ref{firstlaw})],
and the small $Q>0$ is just to keep $T$
constant.  One may check that
eqs. (\ref{dh=q}) (for $H_{\parallel}$)
and (\ref{df=-w}) hold in this isothermal process.

\vspace{0.3cm}

{\it Constant $U$}

\vspace{0.3cm}

Consider a reversible process with $U=U_{0}$ throughout
\footnote{Note that $U$ in eq. (\ref{ltu}) is negative.
}.
Before calculating $W$, notice that $P_{\parallel}$ 
in eq. (\ref{ltp}) can be recast as
\begin{equation}
P_{\parallel}=\frac{\pi^{2}}{480}\frac{\hbar c}{d^{4}}
+\frac{U_{0}}{2Ad}.
\label{ltp2}
\end{equation}
Then, noting eq.  (\ref{ltp2}),
integration of  $P_{\parallel}d$ over $A$ yields
\begin{equation}
W=Q=\frac{\pi^{2}}{480}\frac{\hbar c}{d^{3}}\left(A_{f}-A_{i}\right)
+\frac{U_{0}}{2}\ln\left(\frac{A_{f}}{A_{i}}\right).
\label{dltwu}
\end{equation}
A further analysis shows that the first term in eq. (\ref{dltwu})
is the leading contribution at this regime. Therefore $Q>0$ is
associated with an expansion, i.e., $A_{i}<A_{f}$ with no surprises this time.

Keeping $d$ fixed, we let the system go spontaneously from
$A=A_{i}$ to $A=A_{f}$. Since $P_{\parallel}>0$, an expansion happens.
Instead of eq. (\ref{2dt/dd}), we now have that
\begin{equation}
\left(\frac{\partial T}{\partial A}\right)_{U,d}=
\frac{\pi^{3}}{2160\, \zeta(3)}\,\frac{(\hbar c)^{3}}{kAd^{3}(kT)^{2}}
\left[1-\frac{720\, \zeta(3)}{\pi^{3}}\left(\frac{kTd}{\hbar c}\right)^{3}
\right]>0,
\label{3dt/dd}
\end{equation}
i.e., the system warms up, in contrast
with the conventional blackbody free expansion \cite{kel81}.
By using 
eqs. (\ref{ltp}) and (\ref{ltp2}), it follows  that 
\begin{equation}
\left(\frac{\partial P_{\parallel}}{\partial A}\right)_{U,d}=
-\frac{U_{0}}{2A^{2}d}>0, 
\hspace{2.0cm} 
\left(\frac{\partial P_{\perp}}{\partial A}\right)_{U,d}=0.
\label{2dp/dd}
\end{equation}
Thus, in this free expansion,
eq. (\ref{2dp/dd}) 
tells us that 
$P_{\parallel}$
increases and $P_{\perp}$ remains unchanged, also contrasting
with the usual blackbody thermodynamics.
Of course $S$ increases [cf. eq. (\ref{ltfs})].

\vspace{0.3cm}

{\it Constant $S$}

\vspace{0.3cm}

Now eqs. (\ref{ltadiabatic}) and (\ref{ltadiabatic2}) still apply, and
eqs. (\ref{du=-w}), (\ref{ltu}) and (\ref{ltp}) lead to
\begin{equation}
W=d\left[(P_{\parallel})_{f}\, A_{f}-(P_{\parallel})_{i}\, A_{i}
+\frac{3\,\zeta(3)}{2\pi d}
\frac{k^{3}}{(\hbar c)^{2}}
T_{i}^{2}\left(T_{i}-T_{f}\right)A_{i}
\right],
\label{dltws}
\end{equation}
where we have used that $T_{i}^{2}A_{i}=T_{f}^{2}A_{f}$
[see eq. (\ref{ltfs})].
By comparing eqs. (\ref{dltws}) and (\ref{dhtws}) one may
be tempted to guess that the signs of the two first terms
in eq. (\ref{dltws}) are incorrect. That is not the case.
As $P_{\parallel}>0$ [see eq. (\ref{ltp})], $W$ must be positive in a expansion. Thus it follows that eq. (\ref{dltws}) is consistent
since $P_{\parallel}A$
increases when $A$ also increases [see eq. (\ref{ltadiabatic2}) and the text
just before it]. For completeness we mention that $P_{\parallel}$
diminishes as $A$ increases.

\vspace{0.3cm}

{\it Constant $H_{\parallel}$}

\vspace{0.3cm}

The Joule-Thomson coefficient
in eq. (\ref{joule-thomsond1})
is easily  obtain from eq. (\ref{ltp}), resulting
\begin{equation}
\mu_{\parallel}=
\frac{2\pi}{3\,\zeta(3)}\frac{d}{k}\left(\frac{\hbar c}{kT}\right)^{2}>0.
\label{joule-thomsona4}
\end{equation}
Then it follows 
that when $P_{\parallel}$ drops the system cools down. As $\mu_{\parallel}$ in eqs.
(\ref{joule-thomsond3}) and (\ref{joule-thomsond2}) is also positive,
there is no obvious inversion temperature this time.
It is worth remarking that the comments in the text just after eq. (\ref{joule-thomsond2}) also apply to $\mu_{\parallel}$ in eq. 
(\ref{joule-thomsona4}).

\subsection{Carnot cycles}
\label{carnot}
A good way to verify the formulae involved in the thermodynamic
processes just studied is to consider Carnot cycles. We have implemented
various of them, for both regimes, i.e., 
$kTd/\hbar c\gg 1$ and $kTd/\hbar c\ll 1$, for $A$ fixed or $d$ fixed,
and all of them led to the Carnot's efficiency
$\eta=1-(T_{c}/T_{h})$,
\label{eta}
as should be. 

Worth mentioning is the Carnot cycle where $A$ is fixed and
$kTd/\hbar c\ll 1$. As we now know [cf. eq. (\ref{ltfs})], 
in this case the system cannot 
go from one state to the other keeping temperature constant 
and varying entropy. Nevertheless, if the exponential small contributions
that have been neglected are for this matter taken into account, the corresponding Carnot cycle can be successfully performed.

\section{Summary and conclusions}
\label{summary}

When one considers hot electromagnetic radiation in a cavity, usually
the volume $V$ of the cavity is assumed to be arbitrarily large.
In other words, one chooses to investigate the regime where 
$kTV^{1/3}/\hbar c\gg 1$, neglecting any effect due to the presence of the walls of the cavity. This approach leads to the familiar blackbody thermodynamics 
which, although hugely successful, contains 
various quantities that are trivial or ill defined such as the heat capacity
at constant pressure $C_{P}$. In this work we studied the consequences of taking into account
the presence of two squared perfect conducting parallel walls 
of area $A\gg d^{2}$, where $d$ is the distance between them. 
Perhaps the main result is that certain ill defined quantities become
well defined when the presence of the walls is acknowledged.

In order to obtain thermodynamics of the electromagnetic field
in this slab we used the Brown-Maclay $\left<T_\mu{}_\nu\right>$.
Each thermodynamic aspect was thoroughly examined in both regimes
$kTd/\hbar c\gg 1$ and $kTd/\hbar c\ll 1$.
Due to the anisotropy of the rectangular parallelepiped cavity 
(slab), two pressures come about
which play important roles:
$P_{\parallel}$ is the pressure on the small walls of the slab, whereas $P_{\perp}$ is the pressure on its large walls.
Surprisingly we found that all
quantities associated with $P_{\perp}$, such as $C_{P_{\perp}}$,
are well defined due to the presence of Casimir's effect.
Such quantities have experimental relevance, and one may wonder if these findings can
be checked in the laboratory.
At this point a clarifying remark is in order. The pressures $P_{\parallel}$ and $P_{\perp}$ are both as real and physical as the familiar blackbody radiation pressure and thus they are measurable. The same can be said of their related thermodynamic potentials.
In fact, it should be recalled that these thermodynamic pressures are the stress components of the 
Brown-Maclay $\left<T_\mu{}_\nu\right>$.

By observing the criteria for stability we shown that the system is
thermally stable but mechanically unstable. It should be remarked that
this is not a particular feature of radiation in the slab.
For example, it can be shown that 
hot radiation in a conducting 
spherical cavity is indeed mechanically stable; but that is not the
case if the cavity is cubic.

By studying various classic thermodynamic processes, certain counter-intuitive
phenomena were spotted. Worth mentioning here is that, at the regime of low
temperatures (and/or in narrow slabs), electromagnetic radiation in the
slab behaves somewhere between a  rubber band and a (2+1)-dimensional photon gas.

Along the text we tried to show that although electromagnetic radiation
at high temperatures (and/or in wide slabs) behaves radically different
from 
electromagnetic radiation
at low temperatures (and/or in narrow slabs) they are part of the same 
phenomenon where they are linked by the principles of thermodynamics.

It seems there are many directions in which this investigation can be extended.
The study of other configurations and other fields come immediately to mind.
But there are some of more speculative nature. Namely, its hard not to note certain similarities with the thermodynamics of a self-gravitating gas,
or perhaps even with black hole thermodynamics.
Though rather speculative as already mentioned they deserve further investigation.

\vspace{1cm}
\noindent{\bf Acknowledgements} -- 
The
work of E. S. M. Jr. has been partially supported by
``Funda\c{c}\~{a}o de Amparo \`{a} Pesquisa do Estado de Minas Gerais'' (FAPEMIG)
and by ``Coordena\c{c}\~{a}o de Aperfei\c{c}oamento de Pessoal de N\'{\i}vel Superior'' (CAPES). H. S. acknowledges a grant from
``Coordena\c{c}\~{a}o de Aperfei\c{c}oamento de Pessoal de N\'{\i}vel Superior'' (CAPES).

\end{document}